\documentclass{optica-article}

\journal{opticajournal} % for journals or Optica Open

\articletype{Research Article}

\graphicspath{{./img/}}

\begin{document}

\title{Optical  "fingerprints" of dielectric resonators}

\author{A.P. Chetverikova,\authormark{1,*} M.E. Bochkarev,\authormark{1} N.S. Solodovchenko,\authormark{1} K.B. Samusev,\authormark{1,2} and M.F. Limonov\authormark{1,2}}

\address{\authormark{1}School of Physics and Engineering, ITMO University, 191002, St. Petersburg, Russia\\
\authormark{2}Ioffe Institute, St. Petersburg 194021, Russia}

\email{\authormark{*}alina.chetverikova@metalab.ifmo.ru} %% email address is required; see note below about the corresponding author designation

% use {asbstract*} to suppress the copyright line. Copyright information will be added in production

\begin{abstract*} 
The complete picture of the optical properties of resonant structures, along with the frequency, quality factor, and line shape in the scattering spectra, is determined by the electromagnetic field distribution patterns, which are a kind of "fingerprint" of each resonant eigenmode. In this paper, we simultaneously analyze the changes in the spectra and the transformation of the field pattern during the topological transitions from a thin disk to a ring with a gradually increasing thickness and further to a split ring. In addition, we demonstrate characteristic optical fingerprints for well-known interference effects such as bound states in the continuum and Fano resonances.

\end{abstract*}

%%%%%%%%%%%%%%%%%%%%%%%%%%  body  %%%%%%%%%%%%%%%%%%%%%%%%%%
\section{Introduction}
A detailed description of electromagnetic resonances in dielectric structures is essential for both the analysis the optical spectra and the designing of the field-matter interaction. Modern computational and experimental methods make it possible to determine not only the frequency, quality factor and separate lines in the total scattering spectrum, but also to visualize the distribution of the electromagnetic field inside and around the dielectric particle. The field patterns, in particular, the analysis of their transformation as the geometric and dielectric parameters of the resonator change, allows a rigorous understanding of the physics of interference phenomena, such as Fano and Kerker resonances \cite{Solodovchenko2022,Won2019,Kivshar2022} or bound states in the continuum \cite{Hsu2016} in terms of the interplay among different eigenmodes and finally allow the structure to be shaped to achieve a certain electromagnetic response.

The patterns of electromagnetic fields are widely used to analyze the physical properties of a variety of optical resonant objects. Visualization of resonant fields has been successfully used in the study of the optical properties of metamaterials and metasurfaces. An example is Ref. \cite{Tuz2018}, where the optical properties of a planar all-dielectric metamaterial made of a double-periodic lattice whose unit cell consists of a single subwavelength dielectric particle having the form of a disk possessing a penetrating hole are studied. The authors present detailed cross-section patterns of electric and magnetic field distribution which are calculated within the unit cell at the different resonant frequencies of Mie-type and trapped modes. To offer an intuitive understanding of the scattered field from silicon dimers under various excitation conditions and gap sizes, the authors show patterns of hybridization of electrical and magnetic modes of an isolated silicon sphere into dimer modes. The far field scattering from dielectric dimers involve resonant transverse dimer-modes and off-resonance longitudinal dimer-modes \cite{Pascale2019}. The field distribution patterns of resonant modes of a microstrip fed rectangular dielectric antenna calculated by the FDTD method are presented in Ref. \cite{Mohanan2007}. To study the light concentration within subwavelength spatial regions, the visualization of magnetoquasistatic displacement current density modes of split ring resonators was used \cite{Miranda2020}. The electric field pattern makes it possible to understand the functioning of a groove gap waveguide with a complex design \cite{Alos2013}. Based on the electric field pattern, a polarization rotator is investigated that is capable to rotate the polarization and propagation by $90^\circ$ at the resonant frequency \cite{Hayat2020}. Also, when studying hydrogen sensors based on a microring resonator, electric and magnetic field patterns were used \cite{Cicek2017}.

The spatial distribution of the electric and magnetic field components play an important role in the interpretation of complex interference effects in dielectric resonators. An instructive example is the tunable invisibility of a dielectric cylinder that transforms from visible to invisible state and vice versa, demonstrated in \cite{Rybin2015,Rybin2017} without any coating layers. Experiments on measuring the spectral dependence of the scattering efficiency and calculations of magnetic field maps have shown that the total intensity of Mie scattering for waves of a certain polarization vanishes under the condition of Fano resonance at any angle of observation. Another example is the distribution of the electric field amplitude for the Mie-like mode and Fabry-P\'erot-like mode offers intuitive insight into the nontrivial physics of the bound states in the continuum (BIC), which have also been found in dielectric cylinders \cite{Rybin2017a,Bogdanov2019}. It can also be noted that the electromagnetic field distribution patterns resemble the patterns of two-dimensional Laue optical diffraction, which is successfully used in various experiments, including the analysis of the transition from two-dimensional photonic crystals to dielectric metasurfaces \cite{Rybin2016}.

In this paper, we present the results of a study of the scattering spectra and electromagnetic field distribution patterns of eigenmodes in three dielectric objects connected by topological changes in the structure. First, we consider the transition from a thin disk to a narrow thin ring with a gradual increase in the inner hole, which results in the formation of optical ring gallery modes (RGMs). This is followed by an analysis of the transformation of the scattering spectra of the ring with an increase in its height at constant inner and outer radii. An increase in height leads to the appearance of many lines in the scattering spectra, the unambiguous interpretation of which becomes possible only as a result of an analysis of the field patterns corresponding to both narrow and broad lines. Next, we briefly describe the scattering spectra and the corresponding field distribution patterns of the split ring as a function of the light scattering geometry. And, finally, the interference phenomena observed in the scattering spectra of the ring are analyzed. We show that both the bound states in the continuum and the Fano resonance are characterized by unique field distribution patterns that allow for an unambiguous interpretation and are their fingerprints.

\section{Numerical calculations}

Numerical calculations were used to describe the optical properties of dielectric resonators (disk, ring, and split ring, all with a rectangular cross section). All calculations were carried out in the Comsol Multiphysics program, which allows using the optical module to find eigenvalues (resonance frequencies) and eigenfunctions (electromagnetic field distributions), as well as the scattering cross section (SCS) $\sigma$. Since Maxwell's equations are scaled in the absence of dispersion, the defining geometric size (for example, the outer radius $R_{\rm out}$) can be chosen arbitrarily. The aspect ratio for the disk and the split ring was $R_{\rm out}/h=14.38$, and for the unsplit ring it varied from $3.83$ to $14.38$. In calculations, the dielectric constant of the resonators corresponded to the experimental sample $\varepsilon = 43$. To calculate the eigenvalues and eigenfunctions in Cosmol, the ``Eigenfrequency'' mode was used, while the resonator was surrounded by PML (perfect matched layer), and the incident wave was absent. The eigenfunctions of a dielectric ring resonator are classified using azimuthal $(m)$, radial $(r)$, and axial $(z)$ indices. In the case of a topological transition from a disk to a ring with a rectangular cross section, the eigenvalues were calculated with a step of $\Delta (R_{\rm out}/h) = 0.01$. When calculating the scattering cross section $\sigma_{sca}$, the incidence of a plane wave with $\mbox{TE}$ polarization $(k_{\rm x},E_{\rm y},H_{\rm z})$ on a ring resonator surrounded by a PML layer was simulated. Next, the Poynting vector of the scattered wave was integrated over the surface of the remote sphere. The scattering cross section $\sigma$ was normalized to $S = 2R_{\rm out} \, h$.

The eigenfunctions of a dielectric ring resonator are classified according to azimuthal $(m)$, radial $(r)$ and axial $(z)$ indices, which form an ordered triple $(m,r,z)$. In this paper, special attention will be paid to the photonic galleries of the ring, which are defined by radial $(r)$ and axial $(z)$ indices, forming an ordered pair ($r,z$).

\section{Optical modes of the disk, ring and split ring}

\subsection{Optical eigenmodes at topological disk-ring transition: where is the boundary?}

%
%%%%%%%%%%%%%%%%%%%%
\begin{figure}[htbp]
   \centering
   %\fbox{\includegraphics[width=\linewidth]{Fig_01}}
   \includegraphics[width=7cm]{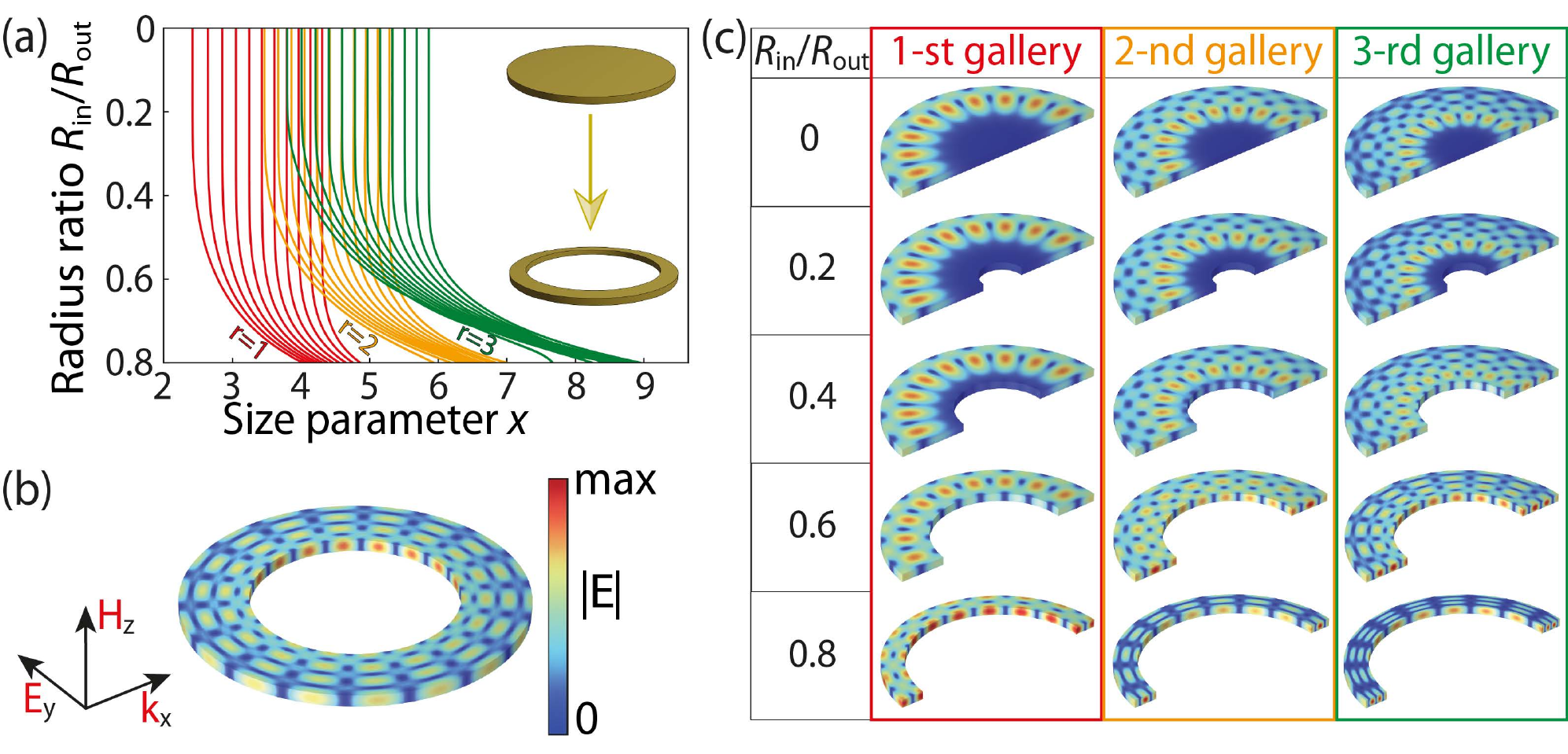}
   \caption
{Topological transition from a dielectric disk to a ring with a rectangular cross section. (a) Dependence of the eigenvalues on the relative value of the inner hole $R_{\rm in}/R_{\rm out}$. The red, yellow, and green lines demonstrate the effect of splitting the initial spectrum of the disk and the formation of three RGMs with radial indices $r=1,\, 2, \,3$, respectively. The inset shows the topological disk-ring transition. (b) A $30^\circ$ view of the electric field amplitude pattern in a resonator with a ratio of radii $R_{\rm in}/R_{\rm out}=0.6$ for the mode with indices $(9,3,0)$ together with the scattering geometry. The vertical color scale corresponds to the amplitude of the electric field $|\mathbf{E}|$. (c) Transformation of patterns of electric field amplitudes distribution for resonances with indices $(9,r,0)$ in the transition from the disk to the ring. The radial index $r$ runs through the values $1, \,2$ and $3$ corresponding to different RGMs. The colors of the columns correspond to the color of the galleries in (a). $R_{\rm in}/R_{\rm out}$ radius ratios are indicated in the left column of the panel (c). All calculations were performed for the permittivity $\varepsilon = 43$ with aspect ratio $R_{\rm out}/h = 14.38$. The resonator is in vacuum, $\varepsilon_{\rm vac} = 1$.
}
\label{fig:fig01}
\end{figure}
%%%%%%%%%%%%%%%%%%%%
%

A change in the resonator topology in passing from a disk to a ring leads to a splitting of the disk spectrum and the formation of a set of individual RGMs \cite{Solodovchenko2022}. Each RGM in the scattering spectrum begins with a broad band associated with a Fabry–Pérot-like transverse resonance and continues with a limited set of narrow equidistant longitudinal modes with an exponentially increasing quality factor. The existence of RGMs was demonstrated in Ref. \cite{Solodovchenko2022}1, but the transformation of eigenfunctions was not studied, and the question of how the transition from the whispering gallery modes (WGM) regime of the disk to the RGM regime of the ring remained open. Recall that WGMs are confined within the structure by total reflections from the rim with the proper phase condition after circling along the whole resonator \cite{Balistreri2001,Righini2011,Baranov2016}.

Figure \ref{fig:fig01} answers this question by showing the transformation of the distribution patterns of the electric field for the first three galleries with radial indices $r=1,\,2, \,3$. The field distribution in the disk $(R_{\rm in}/R_{\rm out}=0)$ for the first gallery corresponds to the classical case of WGM, when most of the resonator volume is not filled, and the field is localized along the outer wall, Fig. \ref{fig:fig01}(c). When the second and third transverse galleries are excited, with increasing index $r$, the electric field in the disk occupies an increasing volume, and with the greatest intensity it is localized not at the outer wall, but shifts closer to the center of the disk [top row in Fig. \ref{fig:fig01}(c)]. Note that when moving to a ring with a small inner hole $(R_{\rm in}/R_{\rm out}=0.2)$, the field patterns practically do not change for all galleries, since in this case only the dark blue part of the disk, which is not filled with an electromagnetic field, is removed [second row of Fig. \ref{fig:fig01}(c)]. However, with a further increase in the inner radius, the electromagnetic field fills the entire ring, the inner wall becomes part of the resonator, and the transverse distribution of the field becomes symmetrical about the center line of the ring. This pattern is typical for the field distribution in the Fabry-P\'erot resonator and corresponds to the RGM regime. Thus, based on the analysis of the electromagnetic field distribution patterns, it is possible to establish the boundary between the WGM and RGM resonant regimes. In particular, in a thin ring with a permittivity $\varepsilon = 43$, the WGM $\to$ RGM transition occurs at a ratio of the inner and outer radii $R_{\rm in}/R_{\rm out} \sim 0.5$, Fig. \ref{fig:fig01}(c).

\subsection{Radial and axial galleries of dielectric ring}

%
%%%%%%%%%%%%%%%%%%%%
\begin{figure}[htbp]
   \centering
   %\fbox{\includegraphics[width=\linewidth]{Fig_02}}
   \includegraphics[width=7cm]{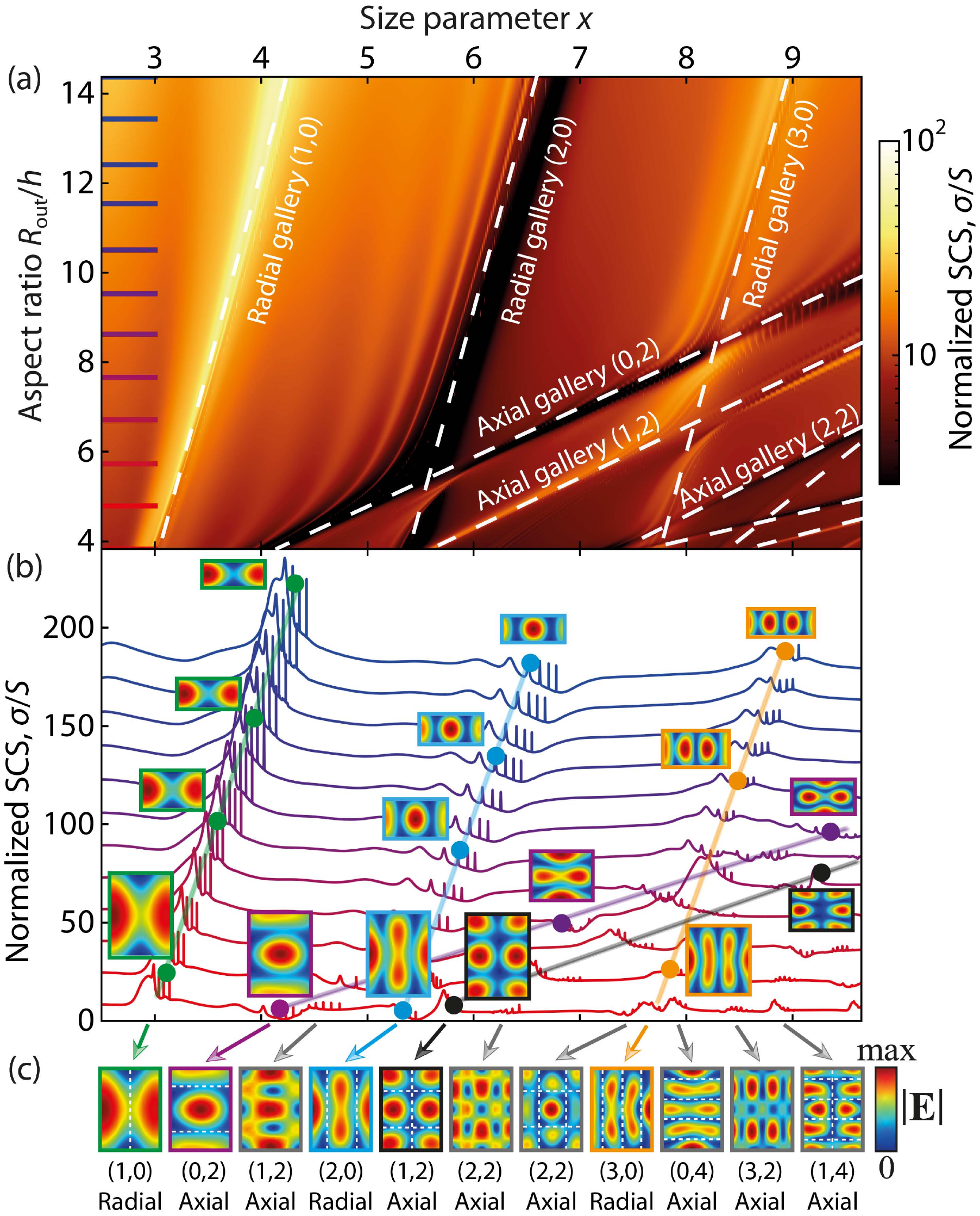}
   \caption
{
(a) Intensity pattern of normalized SCS spectra of dielectric rings for various aspect ratios $R_{\rm out}/h$. The spectra were calculated in the interval $3.83 \le R_{\rm out}/h \le 14.38$, with a step of $\Delta (R_{\rm out}/h) = 0.06$, the total number of spectra $N=167$. The white dashed lines emphasize the linear displacement of the galleries in the scales $(x, R_{\rm out}/h)$. The colored dashes on the left scale mark $12$ values of the $R_{\rm out}/h$ parameter for the spectra shown in panel (b). On the right is a color scale of SCS intensity on a logarithmic scale. (b) Twelve spectra, the color of which corresponds to the color of the dashes on the left in panel (a), marking the corresponding $R_{\rm out}/h$ parameter. The spectra are shifted vertically by a constant value $15$. Above the spectra are the electric field distribution patterns for the selected resonances marked with circles, which are connected by solid lines of the same color. The color of the circles and lines is identical to the color of the frames of the corresponding field patterns. (c) Field distribution patterns for resonances of the lower spectrum. For resonances, a pair of radial $(r)$ and axial $(z)$ indices are indicated that define the gallery, and its type (radial or axial) is also indicated. $\mbox{TE}$ polarization, $R_{\rm in}/R_{\rm out}=0.81$, $\varepsilon = 43$. The normalized size parameter $x = R_{\rm out}\, \omega /c$.
}
\label{fig:fig02}
\end{figure}
%%%%%%%%%%%%%%%%%%%%
%

The central issue that confronts every optician is the correct interpretation of the spectra, which is often a difficult task, especially in the case of a large number of lines that can overlap and interfere. In this section, we will demonstrate how field distribution patterns make it possible to interpret spectra with numerous lines of various widths and shapes.

Figure \ref{fig:fig02}(a) shows the result of a detailed calculation of the light scattering spectra from a dielectric ring depending on its height when an electromagnetic wave is incident in the plane of the ring with $\mbox{TE}$ polarization. The spectra were calculated with a small height step $[\Delta (R_{\rm out}/h) = 0.06]$ in order to obtain an ideal pattern with a high resolution of the resonant features for their further detailed interpretation. In thin rings $(R_{\rm out}/h \ge 10)$ in the discussed spectral range, only three galleries $(1,0)$, $(2,0)$, $(3,0)$ are observed, which begin with a radial Fabry-P\'erot-like resonances between the outer and inner walls of the ring \cite{Solodovchenko2022}. These galleries have a weak monotonic dependence on the height of the ring $h$. New galleries, which begin with an axial Fabry-P\'erot-like resonances between parallel lower and upper walls of the ring appear in the calculated spectra as the ring height increases, when half the wavelength becomes a multiple of the increasing height. These axial galleries, starting with $(0,2)$, $(1,2)$, $(2,2)$ and further, can be immediately distinguished from radial ones by a much stronger dependence on the height h of the ring. Figure \ref{fig:fig02}(a) shows the linear dependences of the axial galleries, which is explained by the choice of scales, namely $x \sim R_{\rm out}/ \lambda$ and $y \sim R_{\rm out}/h$, which is equivalent to the space with $x \sim \lambda$ and $y \sim h$, in which a linear dependence of the Fabry-P\'erot resonance wavelength on the resonator length $(h)$ should be observed.

Thanks to the contrasting logarithmic color intensity scale, Fig. \ref{fig:fig02}(a) makes it possible to establish a number of important features of the SCS spectra of the dielectric ring. As shown earlier \cite{Solodovchenko2022}, the contours of intense radial Fabry-P\'erot-like resonances, which determine the low-frequency SCS spectrum of the ring, alternate in the sequence Lorentz - Fano - Lorentz - Fano, etc. Indeed, rather bright bands are observed on the intensity pattern in the region of the first $(1,0)$ and third $(3,0)$ radial galleries, which indicate intense close to symmetrical Lorentzian lines in the scattering spectra. At the same time, in the region of the second radial gallery $(2,0)$ one observes a transition from a bright region to a wide, almost black region, which corresponds to an asymmetric Fano contour that touches zero at a certain point in the spectrum. It can also be said with certainty that the gallery $(0,2)$ begins with an axial Fabry-P\'erot-like resonance with the Fano contour and, probably, the axial gallery $(2,2)$ has the same character.

In addition to observing Fano resonances, the intensity patterns in Figs. \ref{fig:fig02}(a) demonstrate a number of other interesting features of the scattering spectra. The most striking optical effect is demonstrated by the Fano-galleries $(2,0)$ and $(0,2)$ in the region of the normalized frequency $x=6$, where the galleries demonstrate a clearly pronounced anticrossing, Fig. \ref{fig:fig02}(a). In the low-frequency region of the spectra, the "black" band of the gallery $(2,0)$ continuously passes into the "black" band of the gallery $(0,2)$, and in the high-frequency region, a "yellow" resonance is formed that locally intersects the "black" extension of the gallery $(2,0)$. In addition, the radial gallery $(3,0)$ also does not intersect with the axial galleries $(0,2)$, $(1,2)$, $(2,2)$, showing a series of successive anti-intersections along the dashed line, which allows us to trace the total shift of this gallery in parametric space. Such phenomena in scattering spectra can lead to the formation of BIC \cite{Hsu2016,Doeleman2018,Koshelev2019}, in particular, in accordance with the Friedrich-Wintgen mechanism \cite{Friedrich1985}. In this case, one of the interacting modes sharply narrows and its quality factor increases many times, while the second mode, on the contrary, broadens significantly. If the quality factor of the first mode for a number of reasons does not reach the theoretical limit of infinity, then such states are called quasi-BIC. Quasi-BIC was observed earlier in the spectra of dielectric ring resonators \cite{Rybin2017,Bogdanov2019}. The dependence of the quasi-BIC on material losses and its nontrivial manifestation in the field distribution patterns will be considered below in Section 4A.

Figure \ref{fig:fig02}(b) shows $12$ spectra of dielectric rings with a monotonically varying height $h$. The position of the spectra corresponding to them is indicated by horizontal dashes of the same color on the left vertical scale of the panel (a). The upper spectrum, which corresponds to the height of one of the experimental samples $h = 4 \, \mathrm{mm}$ (the outer and inner radii of the sample, respectively, $57.6$ and $46.5 \, \mathrm{mm}$), was interpreted in detail in Ref. \cite{Solodovchenko2022}1. It should be noted that the intensity chart of the spectra in Fig. \ref{fig:fig02}(a), as well as the set of individual spectra in Fig. \ref{fig:fig02}(b), do not allow us to interpret all the lines that appear in the spectra in the high-frequency region with increasing ring height $h$. The key information for the interpretation of the SCS spectra is provided by the field distribution patterns in the resonator cross section. We plotted the calculated fields for the most intense lines on the spectra in Fig. \ref{fig:fig02}(b) by connecting them with colored lines similar to the dashed lines in Fig. \ref{fig:fig02}(a). Row (c) shows the patterns of the modulus of electric fields $|\mathbf{E}|$ for all resonances observed in the lowest spectrum of panel (b) with the largest number of lines. Note that the shape of the rectangle of electric fields on panel (b) reflects the increase in height with the ring width unchanged. These patterns make it possible to uniquely indicate a pair of indices $(r,z)$ for each resonance. The index is determined by the number of lines with zero value (blue color) of $|\mathbf{E}|$ over the entire width or height of the rectangular cross section of the ring. For example, the mode in the black frame clearly has one zero vertical line and two horizontal lines, so this resonance belongs to the gallery (1,2). Its axial type is determined by a strong linear dependence of the frequency on the height of the ring, as follows from Figs. \ref{fig:fig02}(a, b).

In the case that the field distribution pattern for some line in the spectrum of a ring of a certain height h has a complex shape, as, for example, in the case of bound states in the continuum discussed below, it is necessary to trace the transformation of the field for this resonance with a change in height until the line will shift from the interference region and the field distribution will make it possible to uniquely determine the indices $(r,z)$.

Note the observation of two Fabry-P\'erot-like galleries with the same indices $(2,2)$ at different frequencies, Fig. \ref{fig:fig02}(c). In a parallelepiped with a square cross section, two transverse Fabry-P\'erot resonances are degenerate. However, in the ring, the upper and lower walls are flat and parallel to each other, but the inner and outer side walls are curved. As a result, even in a ring with a square cross section, the degeneracy of transverse resonances is removed, their frequencies are split, and radial and axial Fabry-P\'erot-like resonances are observed at close but different frequencies.

\subsection{Split ring: angular dependence}

%
%%%%%%%%%%%%%%%%%%%%
\begin{figure}[htbp]
   \centering
   %\fbox{\includegraphics[width=\linewidth]{Fig_03}}
   \includegraphics[width=7cm]{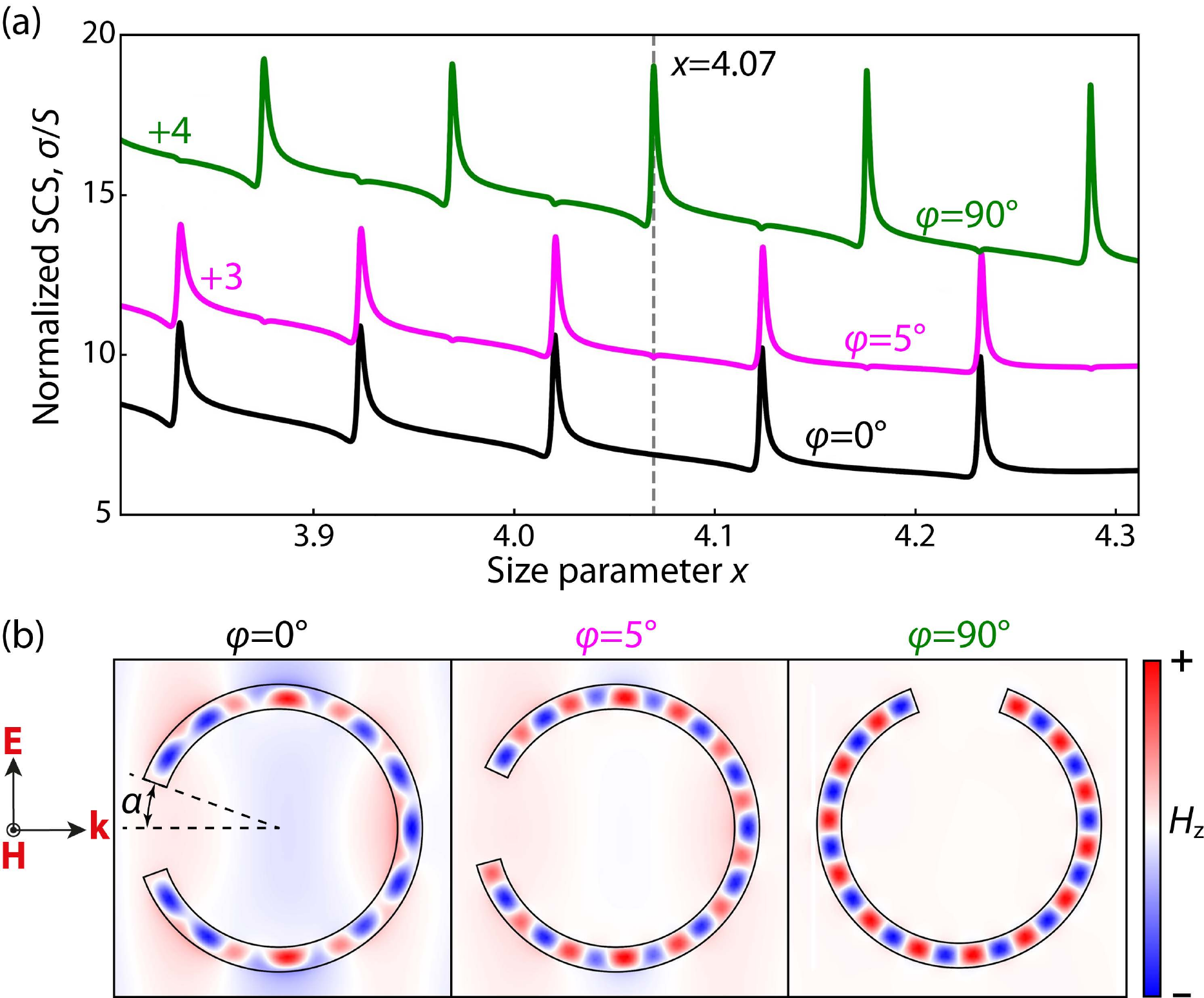}
   \caption
{
(a) Total normalized SCS spectra of dielectric split ring for three different slit orientation angles $ \varphi$ relative to the direction of the incident wave ($X$ axis). The spectra are shifted vertically by the values indicated in the figure on the left. (b) Calculated field distributions in the split ring and around it at the normalized frequency $x=4.07$ marked on panel (a) by a dashed line at the resonant frequency of $3.21  \, \mathrm{GHz}$.  The vertical color scale corresponds to the amplitude of the $z$-component of the magnetic field $H_{\rm z}$. The $Z$ axis is directed along the normal to the ring plane. Ring parameters: $R_{\rm in} = 49.5 \, \mathrm{mm}$, $R_{\rm out} = 60.5 \, \mathrm{mm}$, ring height $h=10 \, \mathrm{mm}$, $\varepsilon = 43$, angle at the center is $2 \alpha=40^ \circ$. The normalized size parameter $x = R_{\rm out}\, \omega /c$.
}
\label{fig:fig03}
\end{figure}
%%%%%%%%%%%%%%%%%%%%
%

%\setcounter{figure}{3}

Split ring resonators have great application potential and the study of their optical properties attracts a lot of attention \cite{Menzel2009,Calandrini2018,RodriguezUlibarri2017}. However, less attention is paid to the study of dielectric split rings, and we fill this gap. Figure \ref{fig:fig03} shows a portion of the normalized SCS spectra of dielectric split ring \cite{Bogdanov2019,Menzel2009,Calandrini2018}  in the region of the first photonic gallery. When a gap is introduced, the azimuthal resonances of the ring transforms into equidistant Fabry-P\'erot-like resonances of $2m$ and $2m+1$ orders ($m$ is an integer, similar to the azimuthal index of the ring). As an example, the resonance of the $m=26$ order at a normalized frequency of $x=4.07$ is considered (corresponds to a frequency of $3.21 \, \mathrm{GHz}$ for a split ring with the parameters indicated in the caption to Fig. \ref{fig:fig03}). Let us define the angle of rotation of the split ring $\varphi$ as the angle of deviation of the bisector of the angle at the center from the direction of the vector $\mathbf{k}$ (horizontal line). Resonances of even orders at the rotation angle $ \varphi = 0^\circ$ are forbidden by symmetry. When a slight asymmetry is introduced into the system by rotation through an angle $ \varphi = 5^\circ$, a resonant state of even order $(m=26)$ begins to form, which is almost imperceptible in the spectrum, but is clearly distinguished in the field distribution pattern by the characteristic periodic pattern of maxima and minima. With an increase in the rotation angle $ \varphi$ to $90^\circ$, a line appears in the scattering spectrum, comparable in width, shape and intensity with the lines of neighboring odd resonances $m = 25, \, 27$, while the contrast of the field distribution in the resonator increases slightly. Thus, the field distribution pattern in the resonator turns out to be more sensitive to the generation of a resonance compared to the scattering spectrum.

\section{Fingerprints of interference effects}

\subsection{Bound states in the continuum: dependence on material losses}

%
%%%%%%%%%%%%%%%%%%%%
\begin{figure}[htbp]
   \centering
   %\fbox{\includegraphics[width=\linewidth]{Fig_04}}
   \includegraphics[width=7cm]{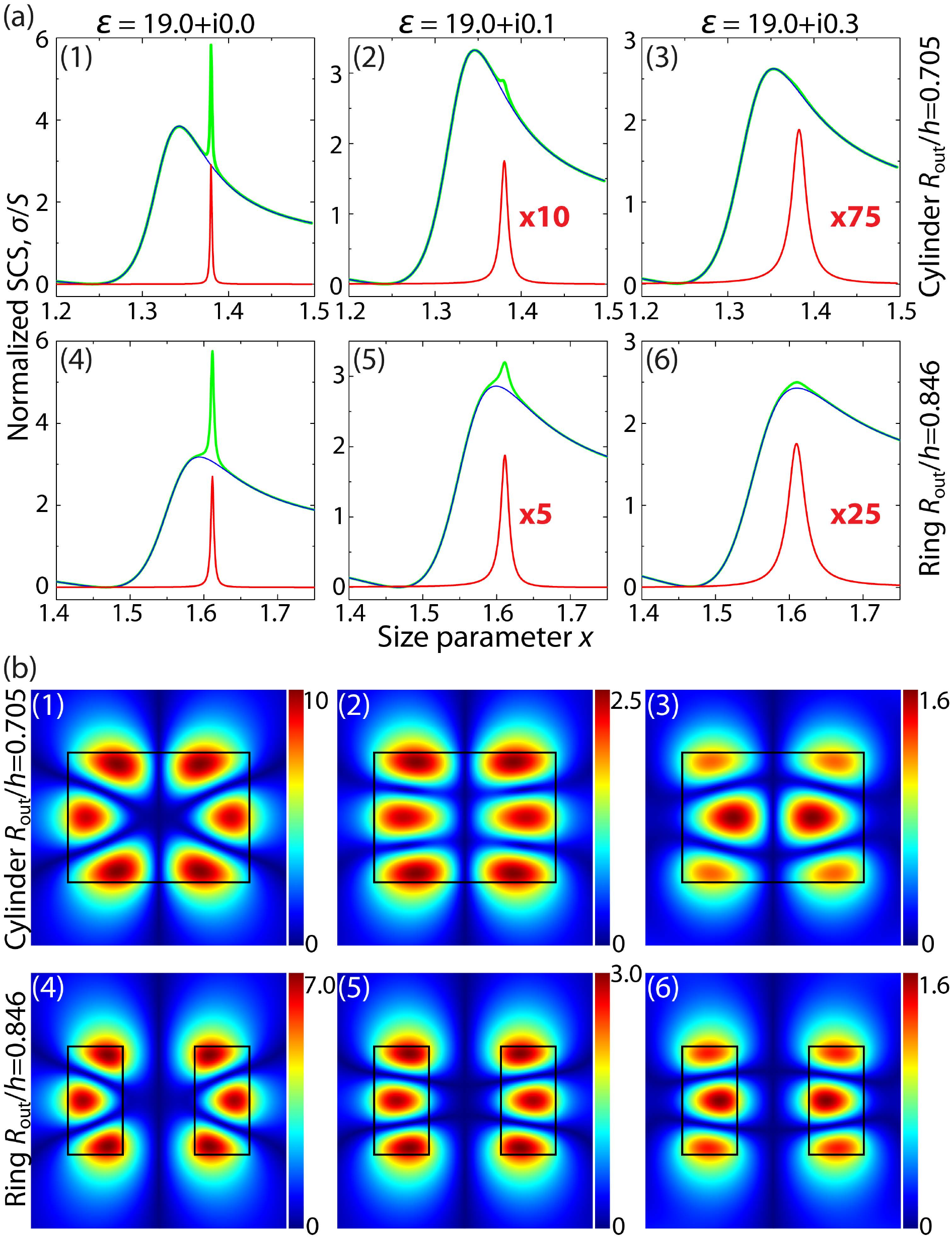}
   \caption
{
(a) Dependence of the SCS spectra of the dielectric cylinder [(a) 1-3)] and the ring [(a) 4-6) corresponding to the quasi-BIC regime (blue line) on the magnitude of material losses, and the decomposition of the spectra into two contours corresponding to the low-frequency (green line) and high-frequency (red line) branch. $\mbox{TE}$-polarized incident wave. Normalized size parameter $x = R_{\rm out}\, \omega /c$. (b) The electric field amplitude $|\mathbf{E}|$ distribution patterns versus material losses. The patterns for the high-frequency branches (resonances with red contours on the left panel) of the cylinder [(b) 1-3)] and the ring [(b) 4-6] are presented for quasi-BIC.
}
\label{fig:fig04}
\end{figure}
%%%%%%%%%%%%%%%%%%%%
%

In Section 3B, we noted the existence of the quasi-BIC regime, which is observed in the scattering spectra of dielectric rings. Similar results were obtained earlier for a dielectric cylinder \cite{Rybin2017a,Bogdanov2019}. In this section, we will demonstrate how the high-frequency mode of an interfering pair of modes, corresponding to the quasi-BIC, changes depending on material losses. Figure \ref{fig:fig04}(a) demonstrates that two strongly overlapping contours are observed in the region where the quasi-BIC occurs, a wide one with a characteristic asymmetric shape of the Fano contour and a narrow one corresponding to the quasi-BIC. Moreover, with an increase in losses, the intensity of the quasi-BIC contour decreases dramatically, therefore, in order to determine the distribution of the field, it is necessary to accurately set the frequency of the weak line, which can only be done by separating the two contours mathematically. Previously, it was demonstrated that all lines in the scattering spectra of dielectric cylinders and rings have the Fano contour, which is due to the interference of the reradiated intrinsic resonance and nonresonant background scattering on the structure \cite{Fano1961,Limonov2017,Limonov2021}. The Fano resonance is described by the universal formula:
%
%
%%%%%%%%%%%%%%%
\begin{equation}
   \sigma(\omega)=D^2\frac{(q+\Omega)^2}{1+{\Omega}^2},
   \label{eq:eq01}  
\end{equation}
%%%%%%%%%%%%%%%
%
%                                                               
where $q=\cot\delta$ is the Fano asymmetry parameter, $D^2=4\sin^2\delta$, $\delta$ is the phase difference between a narrow line and a broad line, $\Omega={2}\left(\omega-\omega_0\right)/\Gamma$ is the dimension-less frequency, $\Gamma$ and $\omega_0$ are the width and frequency of the narrow line (quasi-BIC in our case). To decompose the blue spectra in Fig. \ref{fig:fig04}(a) we used two Fano profiles and a slowly changing background:
%
%%%%%%%%%%%%%%%
\begin{equation}
   \mathrm{Norm. \, SCS}(x)=\mathrm{BG} + \mathrm{A}_1 \frac{(q_1+\Omega_1)^2}{1+{\Omega_1}^2} +  \mathrm{A}_2 \frac{(q_2+\Omega_2)^2}{1+{\Omega_2}^2} ,
   \label{eq:eq02}  
\end{equation}
%%%%%%%%%%%%%%%
%
%
The results of the decomposition of the spectra are shown in Fig. \ref{fig:fig04}(a) for a cylinder and a ring at three values of material loss. The broad low-frequency band hardly changes in amplitude, while the low-frequency peak corresponding to the quasi-BIC decreases in amplitude by approximately two orders of magnitude and broadens significantly. Decomposition of the SCS spectra into two Fano contours made it possible to determine the line frequencies and calculate the field distribution exactly for the quasi-BIC mode. First, the results demonstrate the fingerprint of the BIC: the interference mechanism is such that the high-frequency branch corresponding to the quasi-BIC has a non-trivial field distribution pattern with star-type regions of zero intensity (dark blue stripes between bright spots in Figs. \ref{fig:fig02}, \ref{fig:fig04}), and the number of "rays" of the star is determined by the symmetry of the interfering modes, for example $4$ or $6$. The low-frequency counterpart has the usual field distribution with horizontal and vertical dark blue stripes of zero intensity, although directly involved in the destructive interference. Secondly, with an increase in material losses, the low-frequency peak almost does not change in amplitude, while the quasi-BIC decreases in amplitude by approximately two orders of magnitude and noticeably broaden, Fig. \ref{fig:fig04}(a). The disappearance of the quasi-BIC regime is clearly visible in the field distributions: the star-like patterns disappear, changing to the usual pattern formed by horizontal and vertical stripes of zero intensity, Fig. \ref{fig:fig04}(b).

\subsection{Fingerprint of Fano resonancs}

%
%%%%%%%%%%%%%%%%%%%%
\begin{figure}[htbp]
   \centering
   %\fbox{\includegraphics[width=\linewidth]{Fig_05}}
   \includegraphics[width=7cm]{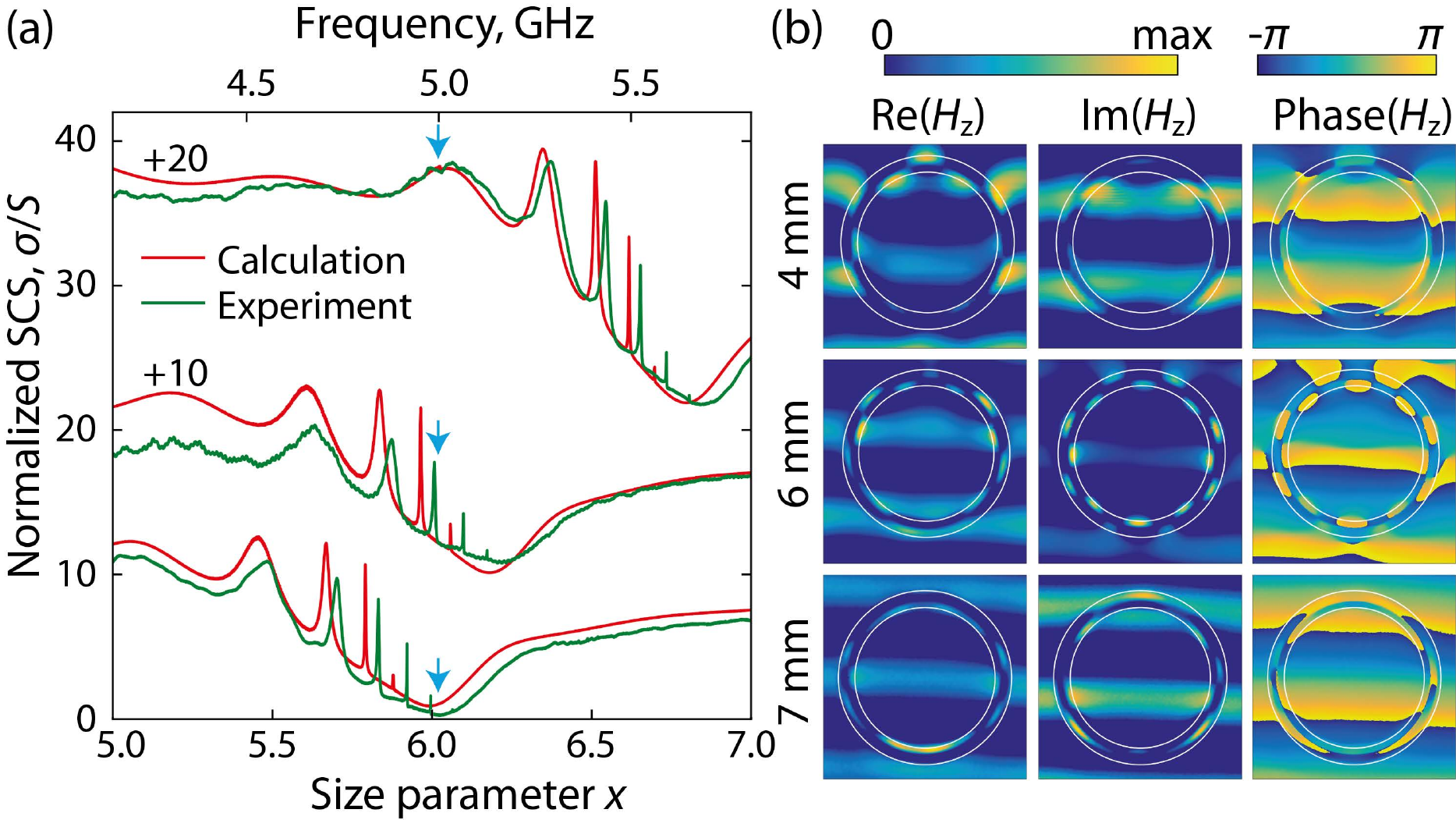}
   \caption
{
(a) Experimentally measured (green lines) and calculated (red lines) far-field normalized SCS spectra of dielectric rings as a function of the ring height h in the Fano resonance region, second gallery. The upper spectrum is for a ring $4 \, \mathrm{mm}$ high, the middle - $6 \, \mathrm{mm}$, the lower - $7 \, \mathrm{mm}$. The spectra are shifted vertically by the indicated value. The blue arrow marks the normalized frequency $x=6$. (b) Experimental spectra in the near-field zone. Shown are: the spatial distribution of the real (left column) and imaginary (middle column) $H_{\rm z}$ magnetic component of the $\mbox{TE}$-polarized electromagnetic field and its phase (right column) for normalized frequency $x=6$, blue arrows in panel (a). $\mbox{TE}$ polarization, $x = R_{\rm out}\, \omega /c$, $\varepsilon = 43$.
}
\label{fig:fig05}
\end{figure}
%%%%%%%%%%%%%%%%%%%%
%

For the experimental study of the SCS spectra in the far- and near-field zone, we used $(\mathrm{Ca_{0.67}La_{0.33}})(\mathrm{Al_{0.33}Ti_{0.67}})\mathrm{O_3}$ ceramic samples, the manufacturing technology of which is described in detail in Ref. \cite{Solodovchenko2022}1. The experiments were carried out in the microwave spectral range, in which ceramics have a dielectric permittivity with the real part $\varepsilon = 43$ and minimal losses of $\tan \delta \sim 10^{-4}$. The experimental observation of high-order longitudinal modes and a reasonable agreement between the experiment and the calculation, in which losses were not taken into account at all, testify to the high quality of the studied ceramic sample. The near-field distribution of the magnetic field of the ring illuminated by a plane wave emitted by a rectangular horn antenna was measured using an automatic field scanner and a Langer EMV-Technik magnetic field probe \cite{Solodovchenko2022}. 

In this work, we were interested in the spectral region of interference effects, which leads to the Fano resonance, and in the form of field distribution patterns in rings in this nontrivial range. Of particular interest is due an important feature of the Fano contour, namely the obligatory vanishing of the scattering intensity $\mathrm{SCS}(\omega_{\rm zero})=0$ under the obvious condition $(q+\Omega)=0$, Eq. \ref{eq:eq01}. Earlier, for the case of scattering by a homogeneous dielectric cylinder, it was found that the partial SCS of each harmonic $\mbox{TE}_m$, being a Mie scattering component, is an infinite cascade of Fano resonances \cite{Rybin2013}. The same regularity is observed in the case of a dielectric ring \cite{Solodovchenko2022}, therefore the total SCS spectrum should be written as the sum of the Fano contours: 
%\smallskip
%
 %
%%%%%%%%%%%%%%%
\begin{equation}
   \mathrm{SCS}(\omega)=\sum_{m} D_{m}^{2} \frac{[q_m+(\omega-\omega_{0m})/(\Gamma_{m}/2)]^2}{1+[(\omega-\omega_{0m})/(\Gamma_{m}/2)]^2}.
   \label{eq:eq03}  
\end{equation}
%%%%%%%%%%%%%%%
%
%\\
%
Due to the periodicity law of azimuthal Fano-harmonics $\mbox{TE}_m$ their intensities turn to zero or close to zero at the same frequencies $\omega_{\rm zero}$ [$x \sim 6$ in Fig. \ref{fig:fig05}(a)] that is, for different indices $m$, the following equality holds: 
%\smallskip
%
%
%%%%%%%%%%%%%%%
\begin{equation}
   q_m+(\omega_{\rm zero}-\omega_{0m})/(\Gamma_{m}/2) \approx 0
   \label{eq:eq04} 
\end{equation}
%%%%%%%%%%%%%%%
%
%\\
%
Expression \ref{eq:eq04} indicates the possibility of creating the so-called invisibility regime \cite{Rybin2015} of a homogeneous dielectric ring. Earlier, it was proposed to use plasmonic coatings to mask small dielectric subwavelength particles \cite{Alu2005}. The effect is based on resonant compensation of the dipole moment of a particle if the polarization vector in the plasmonic shell is antiparallel to the polarization vector in the dielectric for a given wavelength. In the Fano resonance regime, the invisibility effect can be obtained without additional coating due to interference cancellation of light scattering, which was first demonstrated using a dielectric cylinder as an example \cite{Rybin2015}. 
Figure \ref{fig:fig05}(b) shows the results of near-field experiments at a fixed frequency of $5 \, \mathrm{GHz}$ for three samples of different thicknesses, and as a result, in three different characteristic regions of the SCS spectrum. At a ring height of $4 \, \mathrm{mm}$, a frequency of $5 \, \mathrm{GHz}$ corresponds to the region of a broad transverse Fabry-P\'erot-like resonance without pronounced axial modes, which fully corresponds to the weakly structured distribution of near field in Fig. \ref{fig:fig05}(b). At a thickness of $6 \, \mathrm{mm}$, a frequency of $5 \, \mathrm{GHz}$ corresponds to the experimentally observed narrow resonance line. The middle row in Fig. \ref{fig:fig05}(b) clearly demonstrates that this is an azimuthal resonance with index $m=8$. For a sample $7 \, \mathrm{mm}$ thick, a frequency of $5 \, \mathrm{GHz}$ corresponds to the characteristic frequency $\omega_{\rm zero}$ of the Fano resonance, and in the far-field zone the scattering intensity of the electromagnetic wave is practically zero. The results in the near-field zone indicate that at frequency of $5 \, \mathrm{GHz}$ neither the amplitude nor the phase of the $H_{\rm z}$ component of the $\mbox{TE}$-polarized field are not distorted, the electromagnetic wave passes through the ring unchanged, remaining an ordinary plane wave [Fig. \ref{fig:fig05}(b), bottom row]. This means that the unchanging field distributions pattern around and inside the ring are the fingerprint of such an interference phenomenon as the Fano resonance.

\section{Conclusions}

In this work, we have studied the scattering spectra and field distribution patterns of three types of dielectric resonant structures coupled by topological transitions - disk, ring and split ring. We have demonstrated that changes in the topology of objects lead to qualitative changes in the spectra: when passing from a ring to a disk, the optical modes break up into separate galleries, and in the spectra of a split ring, the dependence of azimuthal resonances on the mutual position of the slit and the wave vector of the incident wave arises. At the same time, we have demonstrated that without an analysis of the field distribution patterns in the resonator, the correct interpretation of the results and the main conclusions would be either ambiguous or practically impossible. In particular, only on the basis of an analysis of the field distribution patterns can one discuss the transition from the WGM regime in a disk and in a ring with a small hole to the RGM regime in a ring with a sufficiently large inner hole. The field distribution clearly demonstrates the situation when the electromagnetic field fills the entire ring, both side walls of the ring act as resonator walls, and eigenmodes arise associated with the transverse Fabry-Pérot-like resonance in the dielectric ring.
When we call the field distribution patterns as a fingerprint, we mean, first of all, such interference effects that are not trivial for interpretation, such as quasi-BIC or Fano resonance. Directly opposite situations are observed: in the case of quasi-BIC in the regime of anti-crossing of two branches, this is a complex pattern resembling a complex superposition of the fields of two modes, and in the case of a Fano resonance between a narrow and broad resonances, this is, unexpectedly, the absence of any field violation when the wave passes through an object without noticing it. Thus, the analysis of the electromagnetic field distribution patterns is an important tool in resonant optical spectroscopy.

\begin{backmatter}
\bmsection{Funding} 
Ministry of Science and Higher Education of the Russian Federation (Project 075-15-2021-589), Russian Science Foundation (Project No 23-12-00114).

\bmsection{Acknowledgments} 
AC, NS, MB (experiments and calculations) acknowledges the financing of the project No 23-12-00114, KS and ML (calculations and paper preparation) acknowledges the financing of the Project 075-15-2021-589. 

The authors thank E. Nenasheva (Ceramics Co.Ltd., St. Petersburg) for providing samples for measurements.

\bmsection{Disclosures} 
The authors declare no conflicts of interest.

\medskip

\bmsection{Data Availability Statement}
Data underlying the results presented in this paper are not publicly available at this time but may be obtained from the authors upon reasonable request.

\end{backmatter}

\bibliography{Refs}

%%%%%%%%%% If preparing manually:
% \begin{thebibliography}{1}
% \newcommand{\enquote}[1]{``#1''}

% \bibitem{Zhang:14}
% Y.~Zhang, S.~Qiao, L.~Sun, Q.~W. Shi, W.~Huang, L.~Li, and Z.~Yang,
%   \enquote{Photoinduced active terahertz metamaterials with nanostructured
%   vanadium dioxide film deposited by sol-gel method,}
%   {\protect\JournalTitle{Optics Express}} \textbf{22}, 11070--11078 (2014).

% \bibitem{Optica}
% {Optica}, \enquote{{Optica Publishing Group},}
%   \url{http://www.opg.optica.org}.

% \bibitem{FORSTER2007}
% P.~Forster, V.~Ramaswamy, P.~Artaxo, T.~Bernsten, R.~Betts, D.~Fahey,
%   J.~Haywood, J.~Lean, D.~Lowe, G.~Myhre, J.~Nganga, R.~Prinn, G.~Raga,
%   M.~Schulz, and R.~V. Dorland, \enquote{Changes in atmospheric consituents and
%   in radiative forcing,} in \enquote{Climate Change 2007: The Physical Science
%   Basis. Contribution of Working Group 1 to the Fourth assesment report of
%   Intergovernmental Panel on Climate Change,}  S.~Solomon, D.~Qin, M.~Manning,
%   Z.~Chen, M.~Marquis, K.~B. Averyt, M.~Tignor, and H.~L. Miler, eds.
%   (Cambridge University Press, 2007).

% \end{thebibliography}

\end{document}